# Multiple quasi-phase matched resonant radiations induced by modulation instability in dispersion oscillating fibers

Matteo Conforti,[1] Stefano Trillo,[2] Alexandre Kudlinski,[1] Arnaud Mussot,[1],*

[1] *PhLAM/IRCICA, CNRS-Universite Lille 1, UMR 8523/USR 3380, F-59655 Villeneuve d'Ascq, France*
[2] *Dipartimento di Ingegneria, Università di Ferrara, Via Saragat 1, 44122 Ferrara, Italy*
*Corresponding author: arnaud.mussot@univ-lille1.fr*

The propagation of a continuous wave in the average anomalous dispersion region of a dispersion oscillating fiber is investigated numerically and experimentally. We demonstrate that the train of solitons arising from modulation instability is strongly affected by the periodic variations of the fiber dispersion. This leads to the emission of multiple resonant radiations located on both sides of the spectrum. Numerical simulations confirm the experimental results and the position of the resonant radiations is well predicted by means of perturbation theory.

In optical fibers, resonant radiation (RR) results from the perturbation of a fundamental soliton by higher order dispersion when propagating in the low dispersion region [1]. It has been shown that, at first order, the RR frequency shift is ruled by a phase matching relation that depends mainly on the fiber dispersion slope [2]. RRs, also termed Cherenkov radiations, can be generated either on the short or the long wavelength sides of the soliton, depending on the sign of the dispersion slope. In the special case of flattened dispersion optical fibers, where the dispersion slope is extremely weak, dispersion terms of order larger than three must be accounted for. Consequently, two RRs are generated on both sides of the soliton [3].

During the last decade, RR generation has been widely investigated in uniform optical fibers in the context of supercontinuum generation, because it seeds the blue side of their spectrum (see Refs. [4] and [5] for complete reviews). More recently, it has been shown numerically [6,7] and experimentally [8–10] that, by using optical fibers whose group velocity dispersion (GVD) is engineered along their propagation axis, several RRs can be generated from a single soliton. Two different configurations can be distinguished. Firstly, by inducing a relatively slow and non-periodic variation of the fiber GVD compared to the soliton length, multiple collisions of the soliton [8] with the first zero dispersion wavelength (ZDW) of the fiber or multiple crossing through the second ZDW of the fiber [9] by the RR itself leads to additional RRs on the same side of the spectrum. Secondly, if the variation is periodic with a short period compared to the soliton length, a completely different behavior occurs. In these dispersion oscillating fibers (DOFs), multiple RRs are parametrically excited due to the periodic variation of the GVD and are localized on both sides of the soliton spectrum [10]. Note that similar observations have been recently achieved in a passive cavity as a result of its periodic boundary conditions [11]. All these investigations have been performed with short laser pulses (~hundred of femtoseconds) in order to excite only one fundamental soliton. By increasing the pulse duration well beyond this limit, the continuous or quasi-continuous wave (CW) regime is reached, leading to a radically different dynamics. In uniform fibers, it is well known that the CW input field is transformed into a train of soliton pulses by modulation instability (MI) [12]. This process has been widely investigated [13], and it has been shown that each soliton propagating in the vicinity of the fiber ZDW is disturbed by higher order dispersion. Consequently, each of these solitons generate its own RR [14] on one side of the spectrum, which are all localized around the same frequency, since they originate from almost identical solitons. The propagation of a CW optical field has also been investigated in DOFs. It has been shown theoretically [15–18] and experimentally [19–23] that the MI process can be induced by the periodicity of the dispersion. This leads to the generation of many symmetric quasi-phase matched side lobes around the pump. To our knowledge, these investigations have been performed in the normal average dispersion region of the DOFs. As a consequence no train of bright solitons can be generated.

In this Letter, we investigate the propagation of a CW in the average anomalous dispersion region of a DOF. We show that, as in uniform fibers, the average dispersion induces the standard MI process, which transforms the CW field into a train of solitons. Then, by propagating inside the DOF, the solitons shed energy to multiple RRs, as it was demonstrated in Ref. [10] for a single excitation.

The experimental setup is schematized in Fig. 1(a). The pump system is made of a CW tunable laser (TL) diode that is sent into an intensity modulator (MOD) in order to shape 2 ns square pulses at 1 MHz repetition rate. They are amplified by two ytterbium-doped fiber amplifiers (YDFAs) at the output of which two successive tunable filters are inserted to remove most of the amplified spontaneous emission (ASE) in excess around the pump. These quasi-CW laser pulses have been launched along one birefringent axis of the DOF, which outer diameter evolution is shown in Fig. 1(b). It has a sine modulation shape with a 5 m-long period and its average ZDW is 1060.4 nm. The pump wavelength has been fixed to 1070 nm in order to operate in the average anomalous dispersion region (9 nm above the average ZDW of the fiber). We first investigate the MI process with a relatively weak pump power ($P$ = 5 W), in order to work in the undepleted (linear) regime of the MI process. The red curve in Fig. 2(a) represents the experimental output spectrum.

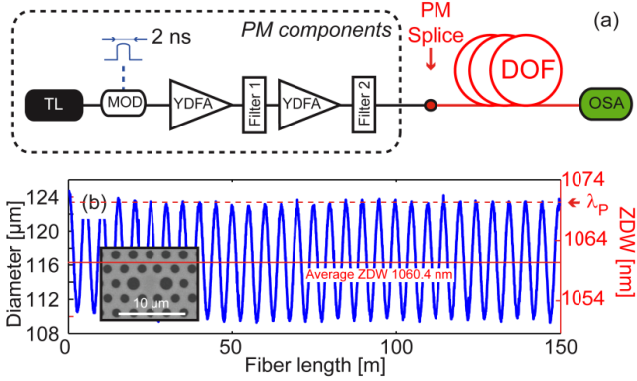

Figure 1 : (a) Scheme of the experimental setup and (b) longitudinal evolution of the outer diameter. Inset: scanning electron microscope image of the fiber input.

Several symmetric side lobes are generated around the pump. Their positions can be predicted by the following quasi-phase matching (QPM) relation that characterizes the MI process in DOFs [15,19]:

$$\overline{\beta_2}\,\Delta\omega^2 + \overline{\beta_4}\,\Delta\omega^4/12 + 2\gamma P = 2k\pi/Z \qquad (1)$$

with $\overline{\beta_{2,4}}$ the second and fourth average dispersion terms, $\Delta\omega$ the angular frequency shift, $Z$ the period of modulation, $\gamma$ the nonlinear coefficient, $P$ the pump power and $k$ an integer. The graphical solution of the QPM relation is represented in Fig. 2(b). The roots (obtained at crossings with horizontal lines) correspond to different MI side lobe orders, $k = 0$ being the standard MI process that would occur in uniform fibers and $k \neq 0$ the ones induced by the periodicity. As can be seen in this figure, the position of MI side lobes is accurately predicted by Eq. (1). It is important to note that MI side lobes that correspond to $k = 0$, are much stronger than the ones induced by the periodicity and lead to the generation of harmonics by beating with the pump (Fig. 2(a)). In fact, it has been demonstrated in Ref. [21] that the parametric gain of QPM MI side lobes due to the periodicity ($k \neq 0$) is always weaker than the one that would correspond to uniform fibers ($k = 0$). These experimental results have been compared with numerical solutions of the generalized nonlinear Schrödinger equation (GNLSE) [13]). We assumed that the DOF has a perfect sinusoidal shape defined as : $\beta_2(z) = \overline{\beta_2} + \delta\beta_2 \times \sin(2\pi z/Z)$ and we used the following parameters that correspond to our experiments: $\overline{\beta_2} = -1.22$ ps$^2$/km, $\delta\beta_2 = -1.2$ ps$^2$/km, $\beta_3 = 0.077$ $ps^3/km$, $\beta_4 = -1.1 \times 10^{-4}$ ps$^4$/km, $\gamma = 10$ /W/km, $\alpha = 5$ dB/km.

The simulated output spectrum is shown in blue curve in Fig. 2(a). As can be seen, a very good agreement is obtained with experiments apart from the noise background level that is higher in experiments. We expect that it mainly originates from the ASE of YDFAs generated between the pulses, which is not accounted for in simulations. It lowers the signal to noise ratio of the side bands but does not affect their dynamics. As can be seen, the relative levels of the side bands are almost similar in experiments and in simulations. We would like to point out that this is the first experimental observation of the MI process in the average anomalous dispersion region of DOFs, where standard MI side lobes ($k = 0$) and those due to the periodicity of the GVD ($k \neq 0$) coexist.

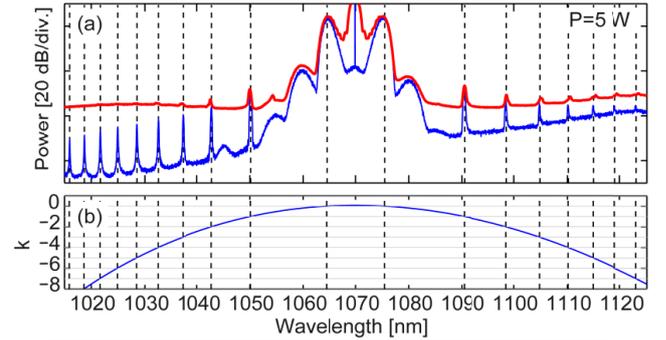

Figure 2: (a) Experimental (red curve) and numerical (blue curve) output spectra for 5 W of pump power. (b) QPM relation. Vertical black lines correspond to the MI side lobe positions determined by the graphical solutions of Eq. (1).

In order to complete our investigations about the dynamics of the process, we gradually increased the pump power from 5 W to 8 W. Corresponding output spectra are displayed in Fig. 3(a). By slightly increasing the pump power to $P = 6$ W, the amplitude of standard MI side lobes increases and additional harmonics appear which overlap with other MI side lobes induced by the periodicity (green curve). Up to this pump power level, all spectral components are perfectly symmetric with regards to the pump as they all originate from four wave mixing processes (phase matched or not). A completely different picture occurs when the pump power is further increased. From 6.5 W of pump power (red and pink curves in Fig. 3(a)), the spectrum keeps broadening and QPM MI sidebands are not visible anymore. Additional sidebands that are asymmetric from the pump appear at large frequency detunings. Their positions do not correspond to QPM processes that appear at lowest pump powers (to be compared with the blue curve, $P = 5$ W). This dynamical evolution versus pump power is confirmed by the numerical simulations reported in Fig. 3(b). There is an excellent agreement with experimental results. Therefore the simulations can be exploited to disclose the dynamics in order to understand this unexpected behavior. The evolution along the fiber of the output spectrum corresponding to the maximum pump power value (8 W, pink curve in Fig. 3(b)) is shown in Fig. 4(a). Up to about 80 m, only the two standard MI side lobes as well as those due to the periodicity are visible. By further propagating inside the fiber until about 120 m length, harmonics of the standard MI side lobes start to appear progressively and overlap the other weaker ones.

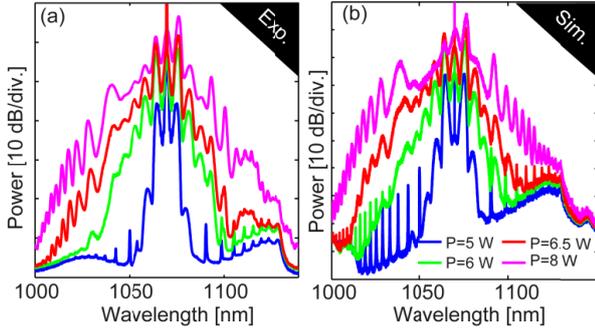

Figure 3: (a) Experimental and (b) numerical output spectra for different pump powers.

The appearance of MI side lobes harmonics is the typical signature of the formation of a train of soliton pulses in the time domain. This is indeed confirmed in Fig. 4(b) where the evolution in the time domain is shown. We see a pulse train with a period of ~0.56 ps, very close to the inverse of the MI frequency shift for $k = 0$ (~0.58 ps, from Fig. 4(a) at $L = 80$ m, $\Delta f \approx 1.7$ THz). From that length, a rapid broadening of the spectrum is observed with the generation of additional side lobes. However, these are not harmonics of the standard ($k = 0$) MI process because they are not symmetric from the pump. Their origin can be explained by applying perturbation theory to the propagation of solitons in DOFs [10]. It has been demonstrated that the propagation of a single soliton is perturbed by the periodic variations of the dispersion. As a consequence it sheds energy to multiple RRs, whose positions are predicted by the following QPM relation :

$$\overline{\beta_2}\Delta\omega^2/2 + \overline{\beta_3}\Delta\omega^3/6 + \overline{\beta_4}\Delta\omega^4/24 - \gamma P_S/2 - \Delta k_1 \Delta\omega = 2m\pi/Z, \quad (2)$$

with $\overline{\beta_3}$ the average value of the dispersion slope, $\Delta k_1$ the deviation of the actual group velocity from the natural one [24,25], $P_S$ the peak power of the soliton when RRs are emitted, and $m$ an integer. In order to facilitate the comparison with the solutions of this equation, the output spectrum of Fig. 4(a) ($L = 150$ m) is represented in Fig. 5(a) (blue curve). Roots of Eq. (2) correspond to the crossing of the dispersion relation with horizontal lines (Fig. 5(b)). Similarly to Eq. (1), $m = 0$ corresponds to RRs emitted also in uniform fibers whereas $m \neq 0$ gives the RRs induced by the periodicity. From numerical simulations, we estimated that the typical values of the peak power of the solitons at which they emit RRs is about $P_S \approx 25$ W and the deviation of the actual group velocity is about $\Delta k_1 \approx 4$ fs/m.

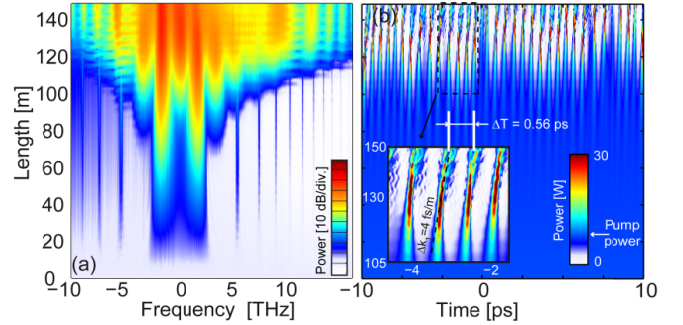

Figure 4: (a) Spectral and (b) temporal evolutions of the input CW field from numerical simulations. Inset : close-up on a few solitons.

As can be seen in Figs 5, the positions of almost all the RRs are very well predicted by Eq. (2) (green lines). For comparison, solutions of Eq. (1) are superimposed in dashed black lines to highlight the fact that this relation cannot predict the positions of RRs. This unambiguously proves that these multiple side bands that appear from $L = 130$ m in Fig. 4 are indeed multiple RRs induced by the periodic variations of the fiber dispersion. The experimental spectrum is superimposed in this figure, and we see that an excellent agreement is achieved with the numerical simulations (red curve).

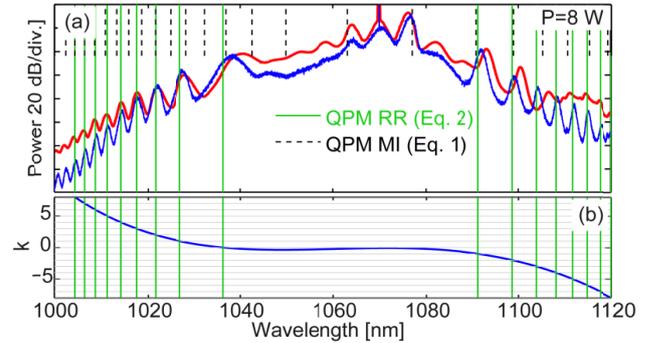

Figure 5: (a) Experimental (red curve) and numerical (blue curve) output spectra for 8 W of pump power. (b) Spectral QPM evolution. The vertical green lines correspond to the RR positions determined by the graphical solution of Eq. (2) while black dashed lines to the one of Eq. (1).

These experimental and numerical investigations demonstrate that the perturbation theory developed in the context of the propagation of a single soliton [10] in DOFs is also valid when a train of solitons propagates inside this fiber. The frequency shifts of RRs generated by solitons is indeed accurately predicted by the relation developed in Ref. [10]. Individual solitons in the train are so close to each other in terms of their parameters that the RRs are generated around the same frequencies. As a result, each side lobe in the spectral domain is indeed composed of many RRs emitted by each soliton. A similar scenario occurs when a CW field propagates in the anomalous dispersion region of an uniform fiber (see Ref. [14] and chapter 8 of Ref. [5] ).

To conclude, we have experimentally and theoretically investigated the propagation of a CW optical field in the average anomalous dispersion of a dispersion oscillating fiber. We have experimentally reported, for the first time to our knowledge, the fact that MI process either induced by the average negative value of the dispersion or by the periodic variation of the dispersion can be observed simultaneously. We have then demonstrated that the standard MI leads to a train of solitonic pulses that are significantly affected by the periodic variation of the dispersion. As a consequence, they shed energy into multiple resonant radiations on both sides of the spectrum whose positions can be accurately predicted by means of perturbation theory [10].

This work was partly supported by the ANR TOPWAVE and FOPAFE projects, by the "Fonds Européen de Développement Economique Régional", by the Labex CEMPI (ANR-11-LABX-0007) and Equipex FLUX (ANR-11-EQPX-0017) through the "Programme Investissements d'Avenir" and by the Italian Ministry of University and Research (MIUR) under Grant PRIN 2012BFNWZ2.